\documentclass[prl,twocolumn,showpacs,footinbib,10pt]{revtex4} %superscriptaddress
\usepackage{bm,braket}
\usepackage{float}
\usepackage[caption=false]{subfig}
\usepackage{graphicx}
\usepackage{lipsum,placeins,afterpage}% http://ctan.org/pkg/lipsum
\usepackage{amsfonts}
\usepackage{amsmath}
\usepackage{amssymb}
\usepackage{color}

\usepackage{hyperref}

\newcommand {\be}{\begin{equation}}
\newcommand {\ee}{\end{equation}}
\newcommand {\bea}{\begin{eqnarray}}
\newcommand {\eea}{\end{eqnarray}}

\renewcommand{\v}[1]{\ensuremath{\mathbf{#1}}} % for vectors

\def\r1{\textbf{r}}

\newcommand{\red}[1]{{\color{black}#1}}

\begin{document}

% change equation skips
\abovedisplayskip=7pt
\abovedisplayshortskip=0pt
\belowdisplayskip=7pt
\belowdisplayshortskip=7pt

\title{Anisotropy-Induced   Quantum Interference and \red{Population Trapping} Between Orthogonal Quantum Dot  Exciton States in  Semiconductor Cavity Systems}
\author{Stephen Hughes$^1$ and Girish Agarwal$^{2,3}$}
\affiliation{$^1$Department of Physics,  Queen's University, Kingston, Ontario, Canada, K7L 3N6\\
$^2$Institute for Quantum Science and Engineering and Department of Biological and Agricultural Engineering, Texas A\&M University, College Station, Texas 77845\\
$^3$Department of Physics,
Oklahoma State University,
Stillwater, OK 74078, USA}
\date{\today}

\begin{abstract} 
\red{We describe how quantum dot  semiconductor cavity systems
can be engineered to realize  anisotropy-induced dipole-dipole coupling between
orthogonal dipole states in a single quantum dot.  Quantum dots in single-mode cavity structures as well as photonic crystal waveguides coupled to
 spin  states or linearly polarized excitons are considered. 
  We demonstrate pronounced dipole-dipole coupling
   to control the radiative decay rate of excitons
  and  form pure entangled states in the long time limit.
We investigate both  field-free entanglement evolution and coherently pumped exciton regimes, and show how a double pumping scenario can completely eliminate the decay of coherent Rabi oscillations and lead to population trapping. 
In the Mollow  regime, we   explore the emitted spectra from the driven  dipoles and show how a non-pumped dipole can take on the form
of a spectral  triplet,  quintuplet, or a singlet, which has applications for producing subnatural linewidth single photons and more easily accessing   regimes
of high-field quantum optics and  cavity-QED.}

\end{abstract}


\maketitle

{\it Introduction.}
The ability to manipulate  spontaneous emission (SE) decay and
coherent coupling between quantum dipoles is a key requirement for
many applications in quantum optics, including the creation of  entangled photon sources and qubit entanglers. Quantum dots (QDs)
  are especially preferable for studying  quantum optical effects due to the large  transition dipole moments. A major problem with entangling
\red{excitons from spatially separated}
  QDs  is due to their large inhomogeneous broadening, leading to negligible photon-coupling rates. 
In 2000, Agarwal~\cite{PhysRevLett.84.5500} showed how  vacuum-induced interference effects
from  an anisotropic  vacuum can lead to quantum interference
effects among decay channels of closely lying states, even though the dipoles are orthogonal: {\em anisotropic vacuum-induced interference} (AVI).
Subsequently, there have been several related theoretical works, though no reported experiments to our knowledge. Li {\it at al.} \cite{PhysRevA.64.013819} demonstrated AVI
using a 3-level atom in a multilayered dielectric medium. Recently,
Jha {\it et al.} \cite{PhysRevLett.115.025501}  studied
a QD coupled to a metamaterial surface to predict AVI using nanoantenna designs, which has the potential advantage of remote distance control; the AVI was shown  to allow a maximum population transfer between the orthogonal dipoles of around 1\%, and similar proposals have been later reported by  Sun and Jiang \cite{OE_MM}. While interesting, these studies are difficult to realize experimentally, and the predicted population transfer coupling effects are rather weak. Moreover,  the plasmonic systems  introduce material losses~\cite{Simon}, and large Purcell factor regimes would be necessary in general.
%Metamaterial systems
%also have very low quantum efficiencies  in general~\cite{Simon}. 
%In addition, all of these proposals consider
%a 3-level system where the decay channels  share the same final state, with %orthogonal spin polarized dipoles.

In practical QD systems, large radiative decay rates
are required and more easily achieved in semiconductor nanophotonic systems. For efficient single photon $\beta$ factors, slow-light PC waveguides have been shown to yield almost perfect single photons on-chip~\cite{RevModPhys.87.347};  such waveguides also exhibit a rich polarization dependence, including points of  linear  and circular polarization.
Charge neutral QD excitons in general exhibit either linear polarization
or circular polarization if the fine structure splitting (FSS) is negligible~\cite{PhysRevB.92.121301,DalacuFSS,MichlerFSS}, which  can  now be controlled with great precision~\cite{TrottalFSS}. Charged QD excitons can also be used to study interactions between
 single spins and photons \cite{EdoSpins}, which is important  for quantum networks.
 It would thus be highly desirable to study and exploit AVI effects in such geometries, using realistic QD
exciton states. Moreover, one would like to go beyond the free-field case of vacuum dynamics and study field-driven coupling via a 
pump field where such effects can  be more easily  accessed and exploited experimentally. \red{Suppressing SE in QDs shows good promise for low error rate quantum logic operations \cite{NatureEdo2014}, and previous attempts to do this are difficult and limited, e.g., using photonic crystal (PC) bandgaps  \cite{PhysRevLett.95.013904}; in addition, the coherent generation of 
 subnatural light from QDs has applications for single photon sources
 \cite{PhysRevLett.108.093602}, and allows one to more easily access interesting strong field physics.  }

\begin{figure}[t]
        \centering\includegraphics[clip,trim=0cm 0cm 0cm 0.cm,width=0.9\columnwidth]{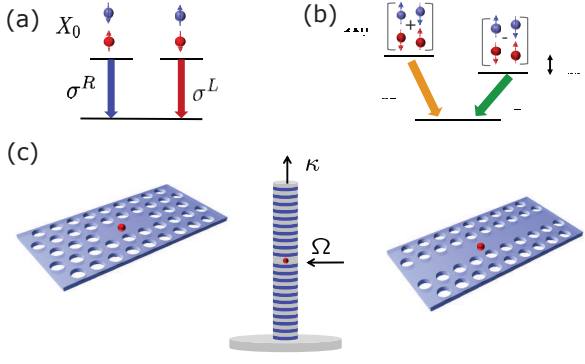}
        \caption{\footnotesize{
       \red{Example QD states, (a) left/right circularly polarized or (b) X/Y linearly polarized, using  neutral  exciton states that can be coupled engineered field modes in a nanophotonic system. The dipole directions are in the plane, caused by stronger quantum confinement in the vertical direction, and the neutral dot excitons may also be split by a small fine structure splitting (FSS).   (c)\ Selection of microcavity and waveguide systems where the field polarization can be controlled, also showing an example of external pumping. }}}
        \label{fig:schematic}
\end{figure}

In this Letter, we introduce several practical, \red{and
experimentally feasible},  QD photonic systems that can  enable \red{and exploit} pronounced AVI, causing long lived entangled photon states with
an almost perfect means of achieving population transfer \red{and population trapping}---a feat that is not possible with spatially separated QD dipoles.
 Figure~\ref{fig:schematic} shows a schematic of   QD exciton states and example photonic systems including a microcavity with a linearly-polarized cavity mode, and a PC\ waveguide that exhibits linear to circular polarization on the so-called L lines (or X points) and C points, respectively
\cite{PhysRevLett.115.153901,Chiral2}.
Such systems provide a high degree of anisotropy needed for observing AVI using QDs. Our significant findings are: ($i$) AVI produces long lived entangled QD states with a population transfer which is orders of magnitude larger than in other systems, \red{ ($ii$) coherent pumping with two pump fields creates a population trapping state in the form of a pure Bell entangled state,  and ($iii$) selective pumping of the transitions enables one to study features of Mollow triplets which are strictly due to AVI, e.g., where one excited dipole acts as the pump for the other dipole and long lives Rabi oscillations can be coherently controlled with great precision.}

{\it Theory.}
Photon transfer can be rigorously modelled through the
the electric-field Green function,
$\mathbf{G}(\mathbf{r},\mathbf{r}';\omega)$
 which describes the field response at  $\mathbf{r}$ to a point source at $\mathbf{r'}$ 
%is defined through
%
%\left [\nabla \times \nabla \times - \frac{\omega^2}{c^2} \epsilon(\mathbf{r})\\ %\right ]\mathbf{G}(\mathbf{r},\mathbf{r}',\omega) =\frac{\omega^2}{c^2}{\mathbf{1}} %\delta(\mathbf{r}-\mathbf{r'}),$
%\label{eq:greens}
%\end{equation}
 where ${\mathbf G}_{i,j}$ is a second rank tensor.
 % and $\mathbf{1}$ is the unit dyad.
% element [$i, j$] corresponds to the response in direction $i$ at $\mathbf{r}$ %from the $j$th component of the source at $\mathbf{r'}$.
Planar PC waveguide modes below the light line ($\omega\!=\!c|\mathbf{k}|$) will propagate without loss through an ideal structure (neglecting imperfections) and can be written as $\mathbf{f}_{k_\omega}(\mathbf{r})=\sqrt{\frac{a}{L}}\mathbf{e}_{k_\omega}(\mathbf{r})e^{i k_\omega x }$, where $\mathbf{e}_{ k_\omega}(\mathbf{r})$ is the Bloch mode, sharing the same periodicity as the lattice, $a$ is the pitch of the PC, and $L$ is the length of the structure; $\mathbf{e}_{ k_\omega}(\mathbf{r})$ is normalized from $\int_{V_c} \epsilon(\mathbf{r})\mathbf{e}_{ k_\omega}(\mathbf{r})\cdot \mathbf{e}^*_{k_\omega}(\mathbf{r'})=\delta_{k_\omega, k_\omega'}$, where $V_c$ is the spatial volume of a PC unit-cell. 
%These normalized Bloch modes can be used to obtain 
The waveguide Green function can be obtained analytically~\cite{Rao2007theory,Yao2009},
\begin{align}
\mathbf{G}_{\rm wg}(\mathbf{r}, \mathbf{r'}; \omega)=& 
\frac{i a \omega}{2 v_g}\Big[\Theta(x-x')\mathbf{e}_{k_\omega}(\mathbf{r})\mathbf{e}^*_{ k_\omega}(\mathbf{r}')e^{i k_\omega (x-x') } \nonumber \\
+&\Theta(x'-x)\mathbf{e}^*_{k_\omega}(\mathbf{r})\mathbf{e}_{ k_\omega}(\mathbf{r}')e^{i k_\omega (x'-x) } \Big],
\label{eq:Gwg}
\end{align}
where the terms preceded by Heaviside functions correspond to forward and backwards propagating modes, respectively,  and $v_g$
%=|v_g(\omega)|$ 
is the group velocity at the frequency on interest. To account for coupling to other modes,
one can simply add other terms to the total Green function,
though these are typically negligible
in comparison to contribution from slow-light Bloch modes.
% :  $\mathbf{G}_{\rm tot} \equiv {\bf G} =\mathbf{G}_{\rm wg}+ \mathbf{G}_{\rm o}$, where the former (dominant) contribution will account for
% coupling to the waveguide mode of interest below the light line, while the latter will account for out of plane decay.
%
%
For a single mode cavity system, with resonant frequency $\omega_c$ and mode
profile, ${\bf f}_{\rm c}({\bf r}),$ the cavity Green function is
\begin{align}
\mathbf{G}_{\rm c}(\mathbf{r}, \mathbf{r'}; \omega)\approx &
\frac{\omega^2 {\bf f}_{\rm c}({\bf r}){\bf f}^*_{\rm c}({\bf r'})}{\omega^2-\omega_{\rm c}^2-i\omega\Gamma_{\rm c}},
\label{eq:G=c}
\end{align}
where at a field antinode the modes can be
normalized through $|{\bf f}_{\rm c}({\bf r}_0)|^2=\eta({\bf r})/V_{\rm eff}\varepsilon_b$,
with $\varepsilon_b$ the background effective index and
$\eta({\bf r})$
accounts for any deviations from the mode antinode position and polarization.

Working in a rotating frame with respect to a laser frequency $\omega_L$,
 we  derive the quantum  master equation (ME) for the  QD interacting with a general photonic reservoir. In the weak-coupling regime, with the system-reservoir coupling given by the dipole interaction in the rotating-wave approximation, we  apply  the second-order Born and Markov approximations to the interaction Hamiltonian, and trace out the photon bath~\cite{PhysRevA.12.1475,Dung2002,PhysRevA.91.051803}.
Thus, the  waveguide and microcavity systems  considered in this work, the coupling rates are  assumed to be in the weak-coupling regime. 
\red{Defining 
$\sigma_{\alpha\beta} = \ket{\alpha}\bra{\beta}$, $\alpha,\beta=g,a,b$, 
   the  ME is
\begin{align} 
\dot{{\rho}}&=i\sum_{n=a,b}{\Delta\omega}_n[  {\sigma}_{nn}, {\rho}]
+i\sum_{n, n'}^{n \neq n'} \delta_{n, n'}[ {\sigma}_{ng}{\sigma}_{gn'}, {\rho}] \nonumber \\ 
&+\sum_{n, n'}\Gamma_{n. n'}\left ({\sigma}_{gn'}{\rho}{\sigma}_{ng}-\frac{1}{2}\{{\sigma}_{ng}{\sigma}_{gn'}, {\rho}\}\right)   
- \frac{i}{\hbar}[{H}_{\rm p}, {\rho}]
\nonumber \\
&+\sum_{n}\gamma'_{n}\mathcal{L}[{\sigma}_{nn}],
%+\sum_{n}\gamma'_{n}\mathcal{L}[{\sigma}^+_{n}{\sigma}^-_{n}],
\label{eq:LME}
\end{align}
where  $n=a,b;n'=a,b$ for two excitons at a QD position ${\bf r}_0$,
$\Delta{\omega}_n =(\omega_L-\omega_n'),$ 
 $\omega_n' = \omega_n -\Delta_{n}$, and  
$\Delta_{n}=\frac{1}{\hbar\epsilon_0}\mathbf{d}_n^\dagger \cdot\text{Re}\left\{ \mathbf{G} ( \mathbf{r}_0, \mathbf{r}_{0};\omega_{n})\right\}\cdot \mathbf{d}_{n}$
is the photonic Lamb shift; $H_{\rm p}  =\sum_{n=a,b}\frac{\hbar\Omega_{n}}{2}({\sigma}_{gn}+ {\sigma}_{ng})$
models a possible external coherent  drive  applied to each dipole,  with an effective Rabi field  $\Omega_{n}=\langle\hat{\mathbf{E}}_{{\rm pump}, n}(\mathbf{r}_n) \cdot\mathbf{d}_n\rangle/\hbar$~\cite{Carmichael1999}; and $\gamma_n'$ models a pure dephasing process. Note that this ME~(\ref{eq:LME}) is more general than the one in \cite{PhysRevLett.84.5500}, 
%as there are no requirements about the dipoles
%decaying to the same final state, 
and we 
also include coupling through the real part of the Green function,  fully accounting
for photon exchange through both real and virtual photons. The dipole-dipole  coupling terms and radiative decay rates are~\footnote{Note ${\rm Re}[{\bf G}({\bf r},{\bf r})]$ formally diverges but we work here with the
transverse Green function (${\bf G}={\bf G}_{\rm wg}$, which has no divergence) and the vacuum contribution is included in the definition of $\omega_{a/b}$.}
\begin{align}
\delta_{n, n'}|_{n \neq n'}=\frac{1}{\hbar\epsilon_0} \text{Re}\left [ \mathbf{d}_n^\dagger \cdot \mathbf{G} ( \mathbf{r}_0, \mathbf{r}_{0};\omega'_{n'})\cdot \mathbf{d}_{n'}\right], 
\label{eq:Lamb_ab}
\\
%\end{align}
%and 
%\begin{align}
\Gamma_{n, n'}= \frac{2}{\hbar\epsilon_0}\text{Im}\left [ \mathbf{d}_n^\dagger\cdot \mathbf{G} ( \mathbf{r}_n, \mathbf{r}_{n}; \omega'_{n'})\cdot 
\mathbf{d}_{n'}\right].
\label{eq:Gamma_ab}
\end{align} 
}

% and QD Rabi fields are chosen to be of similar strength. 
%In the above derivation, the Rabi fields and coupling terms (in units of %frequency) are smaller than the frequency scale over which an appreciable %change in the LDOS occurs, so that the scattering rates are essentially %pump independent~\cite{Ge2013} and the Born and Markov approximations are % valid~\cite{Carmichael1999}.

The usual SE rate from a single dipole in a generalized medium, $\Gamma_a=\Gamma_{a,a}$,
%\begin{align}
%\Gamma_{a}&= \frac{2}{\hbar\epsilon_0}\mathbf{d}_a^\dagger\cdot \text{Im}\left\{ %\mathbf{G} ( \mathbf{r}_0, \mathbf{r}_{0}; \omega'_{a})\right\}\cdot \mathbf{d}_{a},
%\label{eq:Gamma_aa}
%\end{align}
shows that the single dipole emission is proportional to the ${\bf d}_a$-projected LDOS as expected. 
To characterize the strength of the
dipole-medium coupling, we introduce the enhanced  SE\ factor or Purcell factor through
 $F_{\rm P}=\Gamma_{a}/\Gamma^0_{a}$, where $\Gamma^0_{a}$ is the rate for a homogeneous medium.
% which can be obtained by using the homogeneous medium, $\mathbf{G}^{\rm %B}$ in (\ref{eq:Gamma_ab}).
In addition,   there is a possible dipole-dipole coupling term given by
$\Gamma_{a,b}$,
%\begin{equation}
%\Gamma_{a, b}= \frac{2}{\hbar\epsilon_0}\mathbf{d}_a^\dagger\cdot \text{Im}\left\{ %\mathbf{G} ( \mathbf{r}_0, \mathbf{r}_{0}; \omega'_{b})\right\}\cdot \mathbf{d}_{b},
%\end{equation}
and since ${\bf d}_a$ and ${\bf d}_b$ are orthogonal for realistic QDs, this term is usually neglected if ${\bf G}$ is isotropic. However,
as we show below, AVI effects are possible at certain locations, depending upon the
nature of the dipoles and the field modes;
\red{we then exploit such coupling effects to demonstrate a number of striking effects.}
 
\red{
 ($a$) Consider the case of coupled right- and left-CP dipoles,
${\bf d}_a=\frac{1}{\sqrt{2}} ({\bf d}_x +i {\bf d}_y)={\bf d}^R$ and
${\bf d}_b=\frac{1}{{\sqrt{2}}}({\bf d}_x -i {\bf d}_y)={\bf d}^L$,
coupled to a LP field mode, ${\bf E}_k = \alpha{\bf e}^x_k+ \beta{\bf e}^y_k$,
where $\alpha^2+\beta^2=1$. ($i$) If $\alpha =1$,
then $\Gamma_{a,b}=\Gamma_{a,a}=\Gamma_{b,a}$.
($ii$)  If $\beta =1$,
then $\Gamma_{a,b}=-\Gamma_{a,a}=\Gamma_{b,a}$.
($iii$)  If $\alpha =\beta=\frac{1}{\sqrt{2}}$,
then $\delta_{a,b}=\Gamma_{a,a}/2=\delta_{b,a}$.
Remarkably, all three scenarios
can be realized in both cavity and waveguide
systems; indeed, the first two cases can be exploited to 
completely eliminate radiative decay, while the latter case 
is causes by a dipole-dipole induced Lamb shift.
 ($b$) Next,  consider  LP dipoles,
${\bf d}_a={\bf d}_x$
and ${\bf d}_b={\bf d}_y$, coupled 
to an arbitrarily polarized field mode,
${\bf E}_k = \alpha{\bf e}^x_k+ \beta{\bf e}^y_ke^{i\phi}$; here we find that 
dipole-dipole coupling is optimized
when $\alpha=\beta=\frac{1}{\sqrt{2}}$, with
$\phi=0$,
again yielding
$\Gamma_{a,b}=\Gamma_{a,a}=\Gamma_{b,a}$; in this case, clearly
one does not necessarily have to invoke the 
language of an AVI-induced interference, since in this basis the Green function is isotropic.

Note that a C point is rather special; here
 there is no dipole-dipole coupling 
for orthogonal dipoles at the same location; however, generalizing to
the case of two spatially separated dipoles in a waveguide, then one finds a rich variety of dipole-dipole coupling, e.g.,
for RC\ polarized dipoles at two C points,
$\Gamma_{a,b}=2\Gamma_{a,a}\cos[k_\omega(x_a-x_b)]; \Gamma_{b,a}=0,$ [16];  and
for LP dipoles at two C points, then  
$\delta_{a,b}=\Gamma_{a,a}\sin[k_\omega (x_a-x_b)]/2=\delta_{b,a}$.}

{\it Free-field evolution: modified vacuum dynamics.}
Consider  exciton $a$ excited, with 
exciton $b$ in the ground state. For the QD dipoles,
we assume equal resonance energies at $\omega_0/2\pi=200\,$THz
with dipole strength $d=50\,$D, with  
%either ${\bf d}_a = d_x$ ($x$-polarized), ${\bf d}_b=d_y$ ($y$-polarized) %or
${\bf d}_a = {\bf d}_R$  and ${\bf d}_b={\bf d}_L$.
For simplicity we  neglect pure dephasing associated
with charge noise, and  recent experiments~\cite{somaschi} have shown that
such rates can be in the KHz range.
% although our approach can  include such processes,  they do not affect %any of our qualitative predictions that follow.
For the cavity system, we use numbers typical
for PC systems~\cite{Yao2009}, and allow $Q$ to vary,
with $\varepsilon_b=13$ and $V_{\rm eff}=5\times10^{-20}\,{\rm m^{3}}$;
and for the  PC waveguide, we use
$V_{\rm eff}=4\times10^{-20}\,{\rm m^{3}}$,
$n_g=c/v_{g}=50$ \red{(group index)}, $a=400\,$nm. After solving the ME (Eq.~\ref{eq:LME}), the populations
are  obtained from $n_{a/b}(t)=\braket{\sigma_{aa/bb}(t)}$.

\begin{figure}[t!]
        \subfloat[][WG:  $\psi_0=\ket{a}$ at $\frac{1}{\sqrt{2}}$(X+Y)] 
        {\includegraphics[width=0.5\columnwidth]{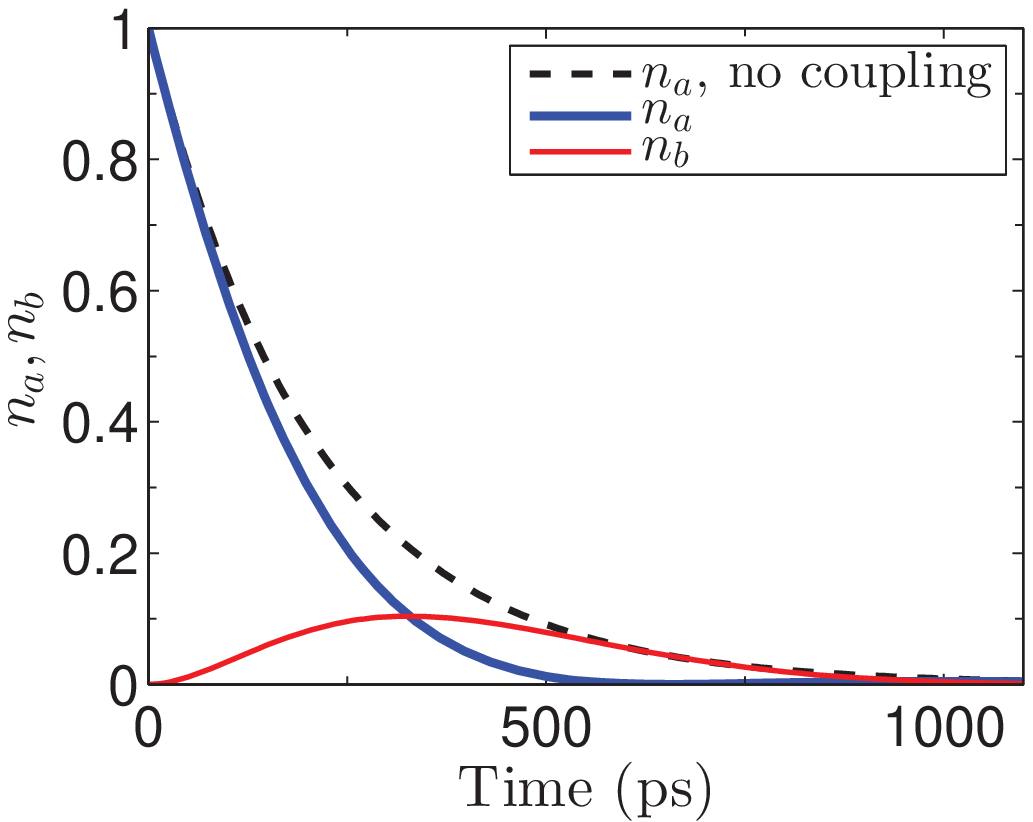}}
        \subfloat[][WG:  $\psi_0=\ket{a}$ at X] 
        {\includegraphics[width=0.5\columnwidth]{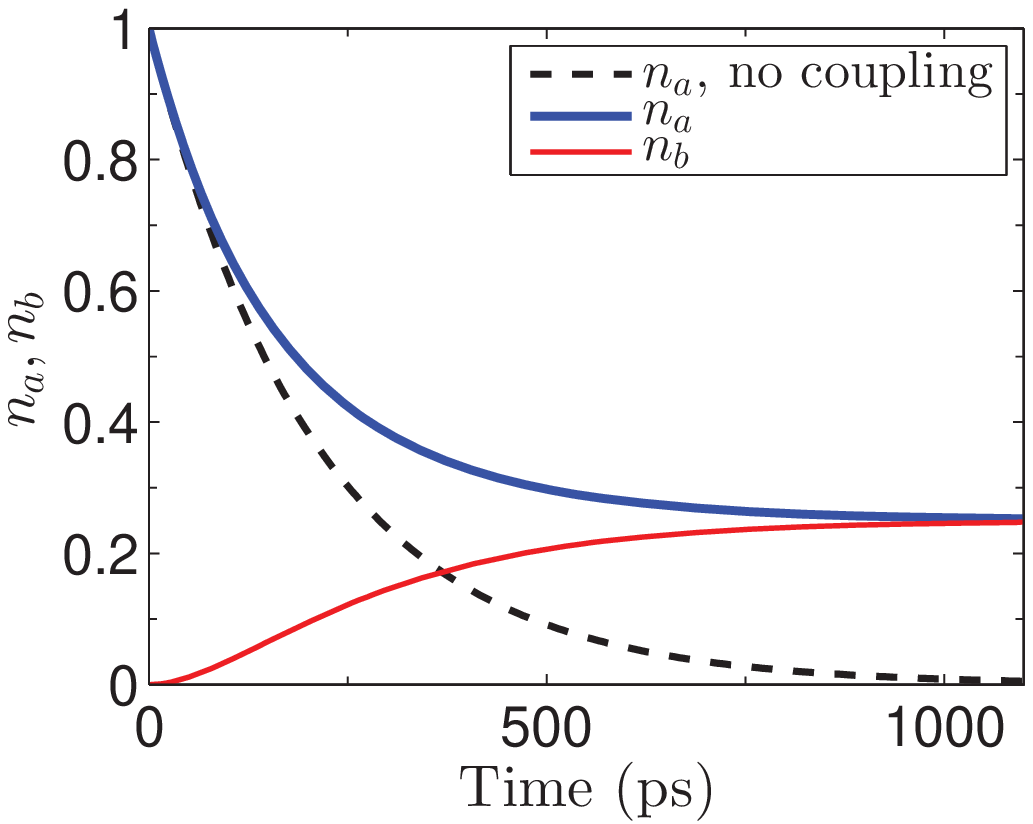}}\\
\vspace{-0.3cm}
         \subfloat[][WG:  $\psi_0=\psi_{+}$ at at X] 
        {\includegraphics[width=0.5\columnwidth]{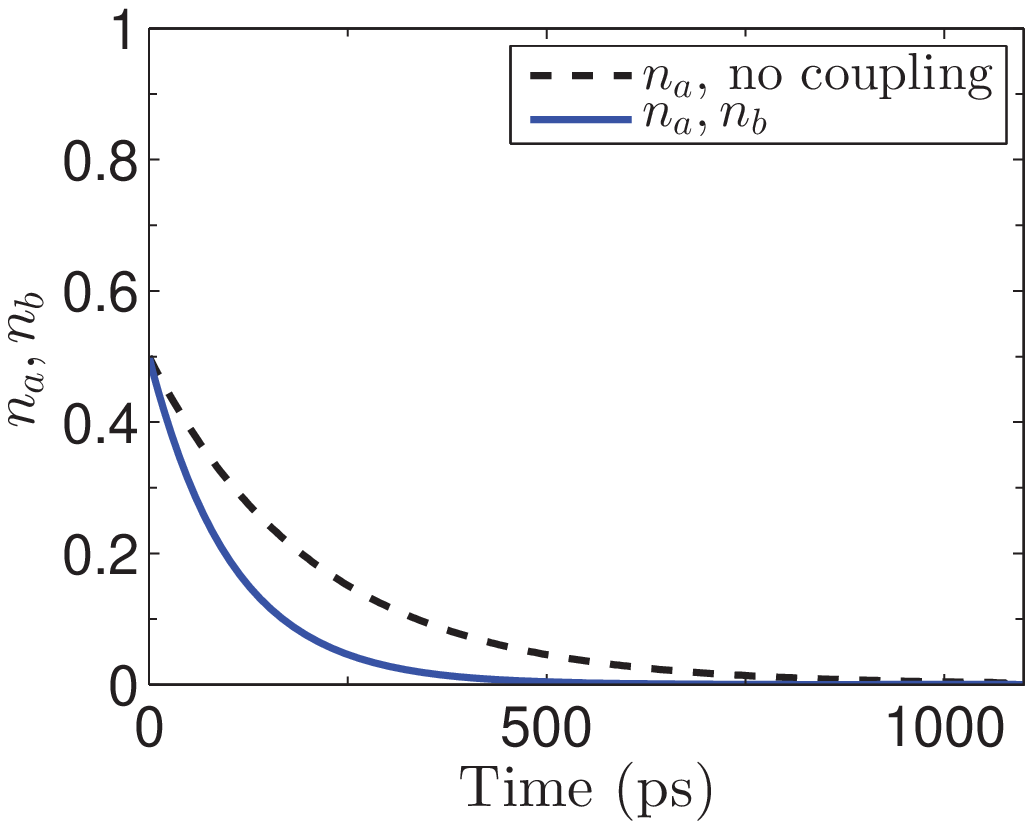}}
        \subfloat[][WG: $\psi_0=\psi_{-}$ at X] 
        {\includegraphics[width=0.5\columnwidth]{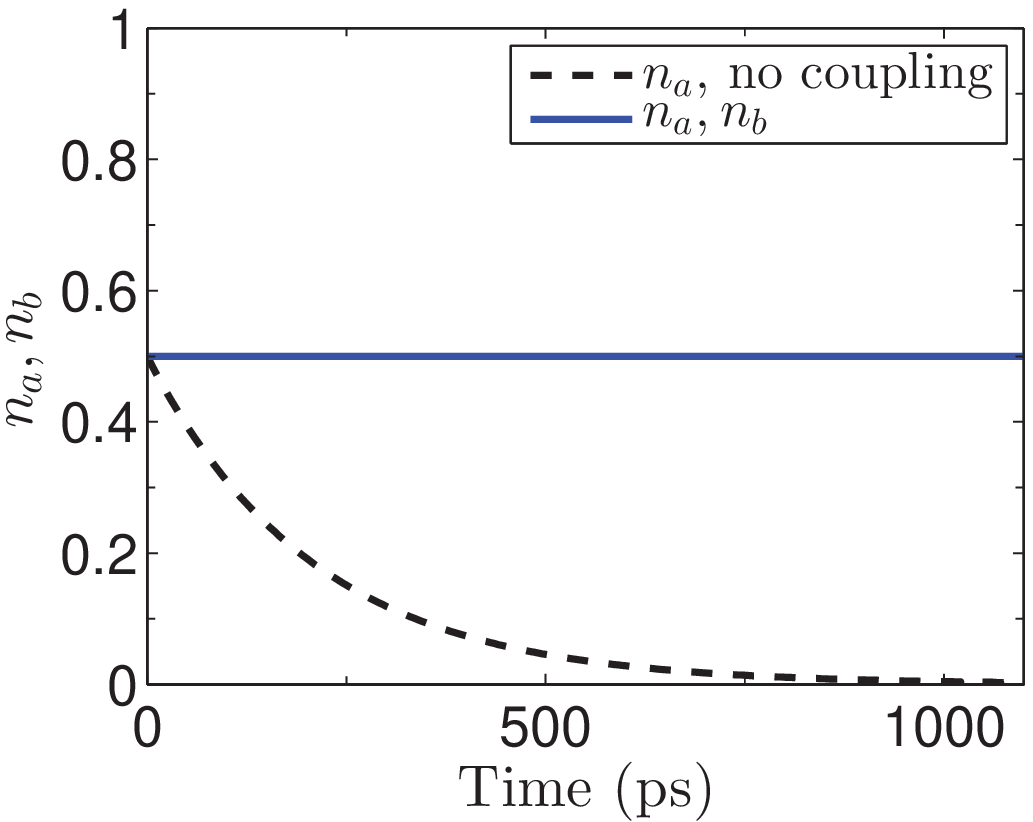}}
        \vspace{-0.1cm}
        \caption{\footnotesize{\red{(a-b) Free field evolution of a  QD two-dipole system with an initially excited QD population inside a slow light waveguide, where (a) is at the $\frac{1}{\sqrt{2}}$(X+Y)] point and (b) is at the X point. Exciton $a$ (blue thick) is initially excited  and the AVI causes QD $b$ (red thin) to be excited.  The population decay without the AVI is shown in black dashed. In case (b), the system forms a pure state consisting of a linear combination of
Bell states $\psi_{\rm -}$ and $\psi_{\rm +}$. (c) Free field evolution with the system in the symmetric state, $\psi_{+}$, showing superradiance. (d) Evolution with an antisymmetric state, $\psi_{-}$, which  stays in a pure excited state in the long time limit.} }}
        \label{fig:free}
        \vspace{-0.1cm}
\end{figure}

\red{Figure~\ref{fig:free}(a) shows the
population dynamics with and without  AVI when $\alpha=\beta=\frac{1}{\sqrt{2}}$
[case $a(iii$)].  We introduce here a new mechanism that to the best of our knowledge is unknown: a Lamb-shift mediated
dipole-dipole interaction between orthogonally polarized excitons, and the amount of population transfer is  quite significant. While $a$ decays faster, exciton
$b$ becomes excited and also decays radiatively.
Next, we consider case $a(i)$ [or case ($b$) with LP\ dipoles].
  The   panels (b-d), show, respectively, the decay
from excited state $a$ excited, and when we start the system
in the antisymmetric and symmetric Bell states:
$\psi_{\pm}=\frac{1}{\sqrt{2}}[\ket{a}\ket{g}\pm\ket{b}\ket{g}]$.
In (b), the system evolves into a linear combination of 
$\psi_{\pm}$, and in (c) we see perfect super-radiance (double the single exciton decay rate); in (d), we completely suppress the radiative decay and evolve into a pure state, with no long lived decay, \red{i.e., an optically dark state}.
For the rest of the paper we consider a QD at the X point, i.e., case $a(i)$.}
 With regards to the corresponding enhanced SE rates,
the Purcell factor in the waveguide,
 $F_{\rm P}\approx 32$; and for the $Q=1000$ cavity, $F_{\rm P}\approx 109$---which are quite modest.

\begin{figure}[t!]
        \subfloat[][WG:  $\Omega_a$ at X] 
        {\includegraphics[width=0.5\columnwidth]{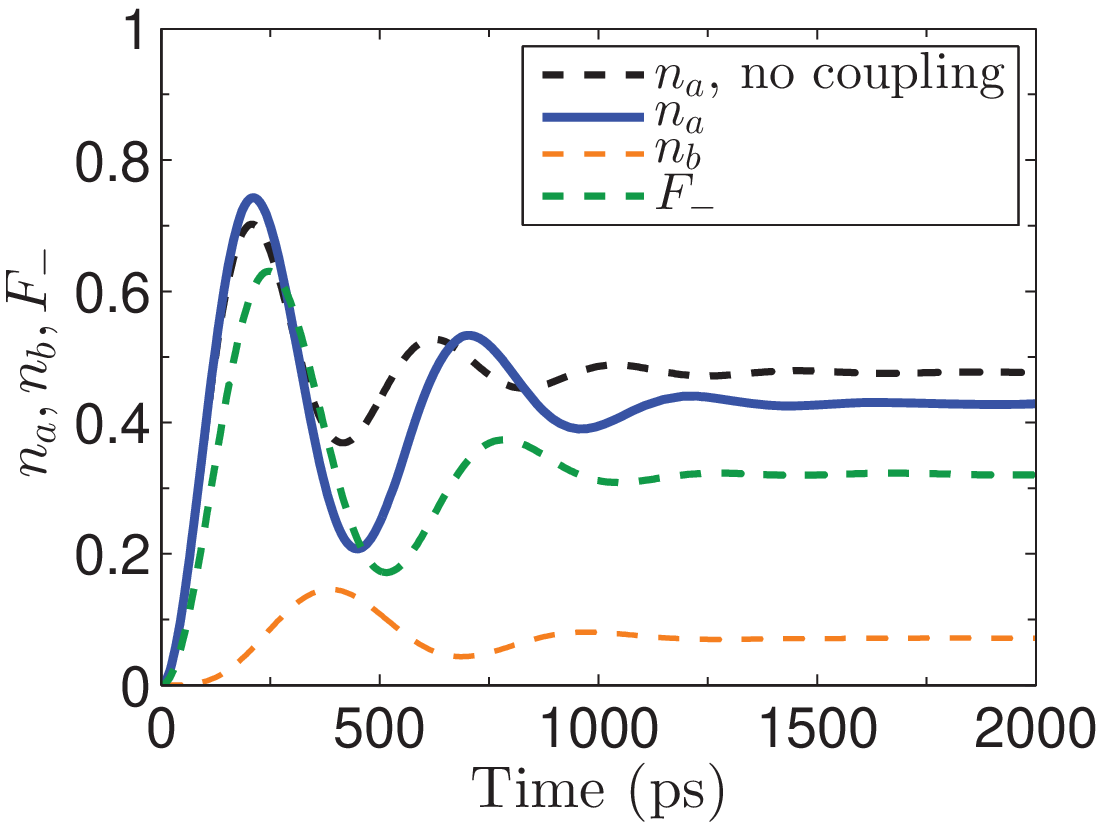}}
        \subfloat[][WG:   $\Omega_{\rm anti}$ at X] 
        {\includegraphics[width=0.5\columnwidth]{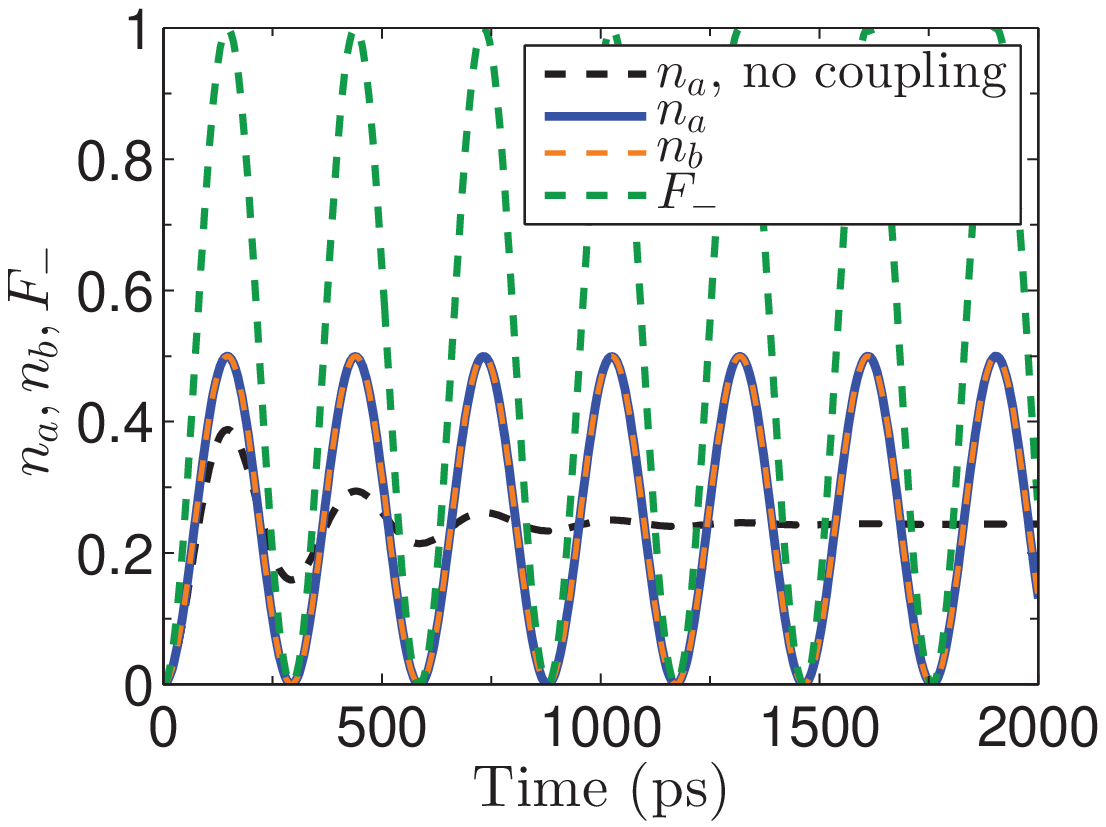}}\\
\vspace{-0.3cm}
         \subfloat[][Cavity:  $\Omega_a$ at X] 
        {\includegraphics[width=0.5\columnwidth]{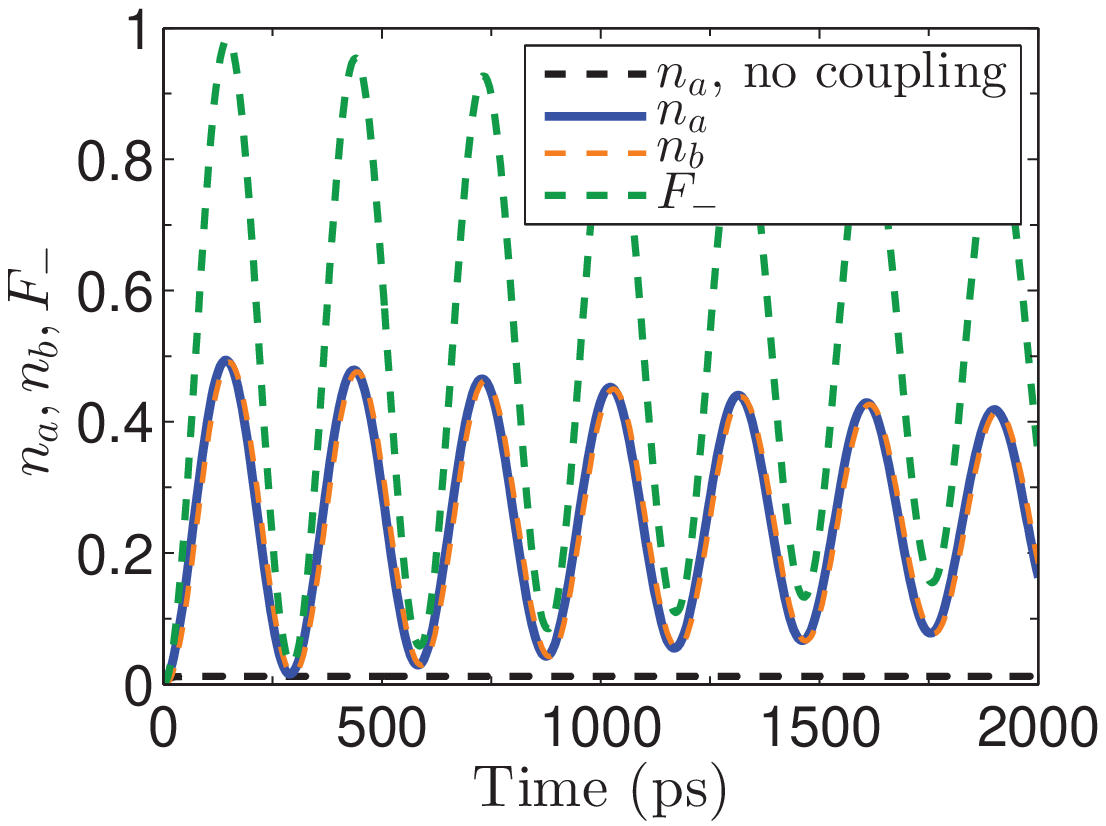}}
        \subfloat[][Cavity:  $\Omega_{\rm anti}$ at X] 
        {\includegraphics[width=0.5\columnwidth]{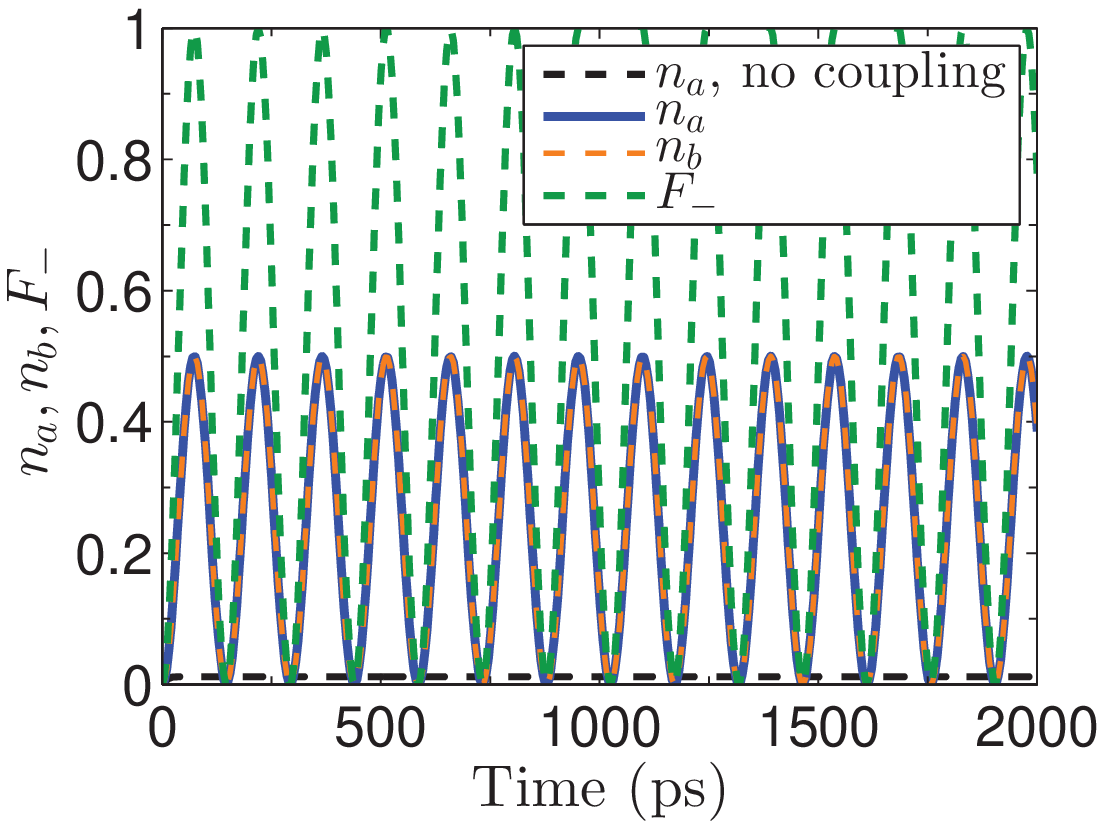}}
        \vspace{-0.1cm}
        \caption{\footnotesize{\red{Examples of a coherently pumped QD two-dipole system. (a) WG (waveguide): Exciton $a$ (blue solid) is pumped with  $\Omega_a=0.01\,$meV  cw driving field, and the AVI causes QD $b$ (red dashed) to be excited. (b) Both excitons are pumped with $\Omega_{a/b}=\pm0.01\,$meV ($\approx 3.7\Gamma_a$). For the cavity (c-d), we use $Q=3000$,
and  $\Omega_0=0.02\,$meV ($\approx 0.06\Gamma_a$). The green dashed curve shows the fidelity of being in the state $\psi{-}$, which clearly exhibits perfect Rabi oscillations with no radiative decay.}}}
        \label{fig:pump}
        \vspace{-0.1cm}
\end{figure}

 {\it CW-pumped entanglement dynamics \red{and population trapping}.}\ 
 Next we look at the situation where one of the QD excitons is coherently pumped, e.g., with an external laser source, and \red{the initial
field  is vacuum with the QD in the ground state}. Normally this would be very difficult to do with spatially coupled dots in the near-field,
% (since both dots would be excited),
but since the dipoles here are orthogonal one can selectively excite only one dipole (or both) with the appropriate pump field polarization. 
\red{In Fig.~\ref{fig:pump}(a), we consider the case where  only exciton $a$ is pumped in a waveguide, which shows good population coupling
and a fidelity to project onto the state $\psi_-$, defined as
$F_-$. In (b), the waveguide system is now excited
antisymmetrically, where $\Omega_a=-\Omega_b$, and this turns out to be the most striking case: we observe the formation of infinite coherent Rabi oscillations, and
a complete suppression of the radiative decay;
in this regime, we have created a population trapping state
which has been studied extensively for multi-level
atom systems ~\cite{GirishBook1,PhysRevLett.81.293,PhysRevLett.77.3995}.
Next we display the $Q=3000$ cavity case in (c-d), and find similar trends, but now even case (c) shows a significant reduction of the radiative decay with only $a$ excited; notably in this case, the
 single exciton case hardly shows any oscillation at all (it is clearly in the weak field regime). Here we see a way to explore high field physics, even though the Rabi field is much smaller than the intrinsic radiative decay rate of a single exciton. Note that
for a $Y$ point [case $a(ii)$], the the trapping solution is simply
$\Omega_a=\Omega_b$, which yields the same trapping state.}

\red{To better explain the creation of a population trapping state,
we have analyzed the optical Bloch equations
from the ME, which reduce to two simple equations:
$\dot\rho_{aa}=i\Omega_02 \rho_{ga}$ and
$\dot \rho_{ga}=i\Omega(1-\rho_{aa})$, with 
$\rho_{ab}=-\rho_{aa}$, 
$\rho_{ga}=-\rho_{gb}$,
$\rho_{bb}=\rho_{aa}$.  These clearly mimic the coherent 
optical Bloch equations for a 2-level atom, but with a factor
of two difference in the  population term. 
Thus, the radiative decay processes cancel by virtue of the AVI-induced coherence, and population trapping occurs.}

 \begin{figure}[t!]
        \subfloat[][$Q$=3000:  $\Omega_a$=$10\,\mu$eV ] 
        {\includegraphics[width=0.5\columnwidth]{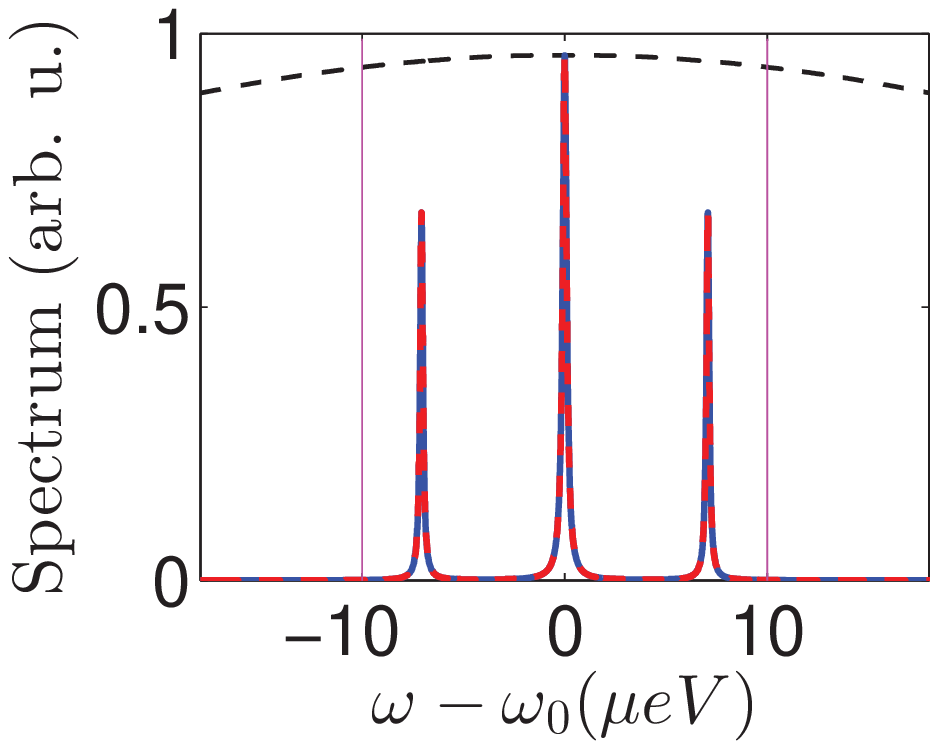}}
        \subfloat[][$Q$=3000:  $\Omega_{a/b}$=$\pm10\,\mu$eV ] 
        {\includegraphics[width=0.485\columnwidth]{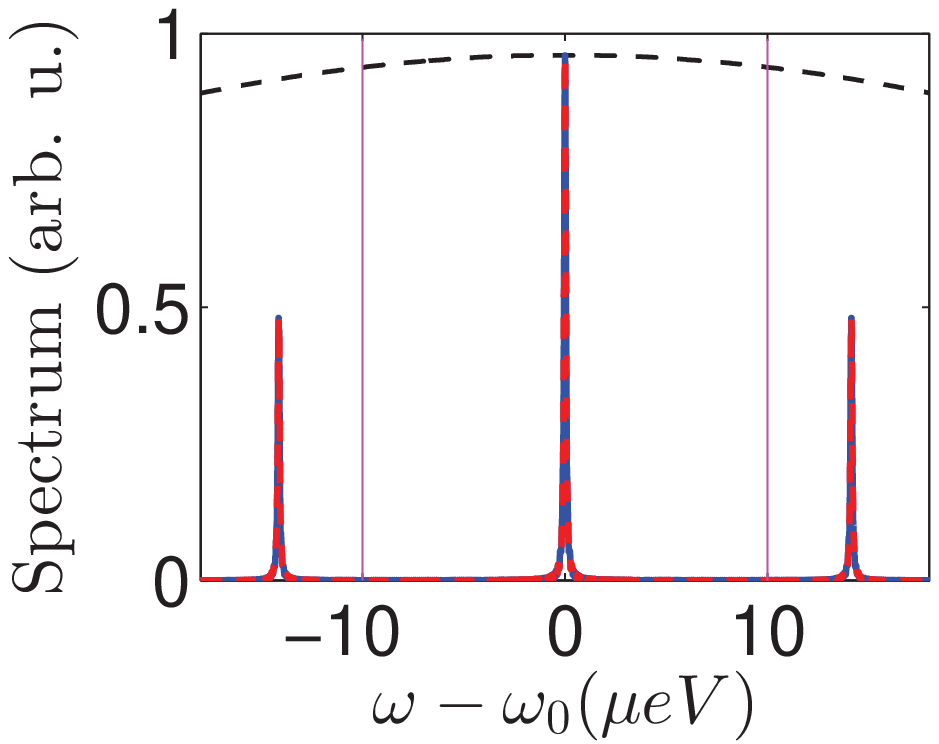}}\\
\vspace{-0.3cm}
         \subfloat[][$Q$=500:  $\Omega_a$=$180\,\mu$eV ] 
        {\includegraphics[width=0.5\columnwidth]{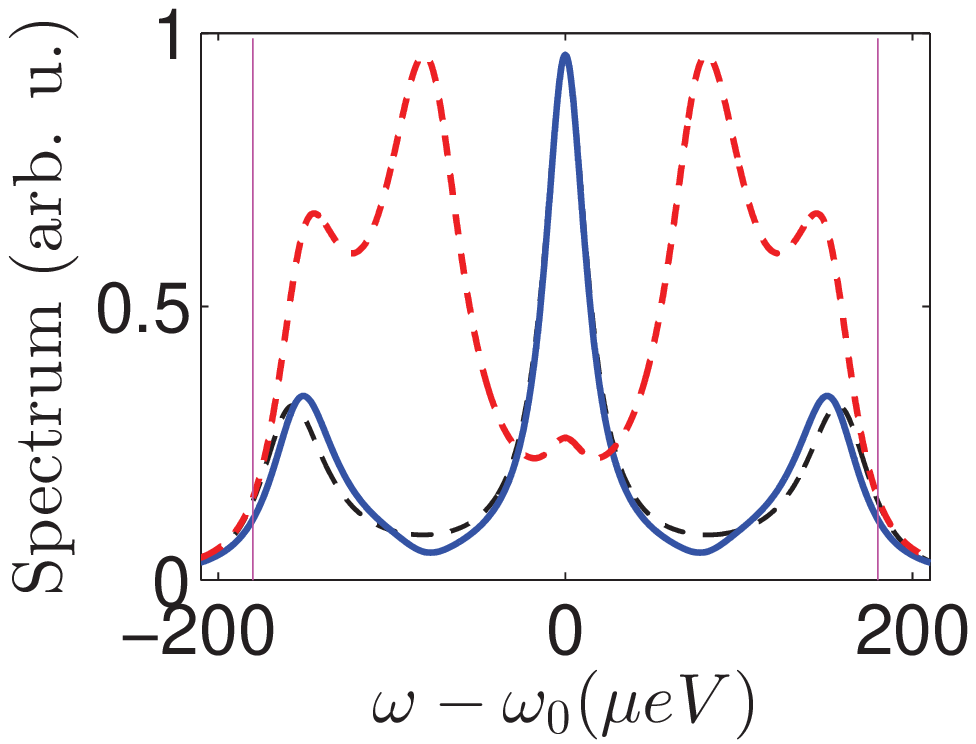}}
        \subfloat[][$Q$=3000:  $\Omega_a=0.2\,\mu$eV ] 
        {\includegraphics[width=0.485\columnwidth]{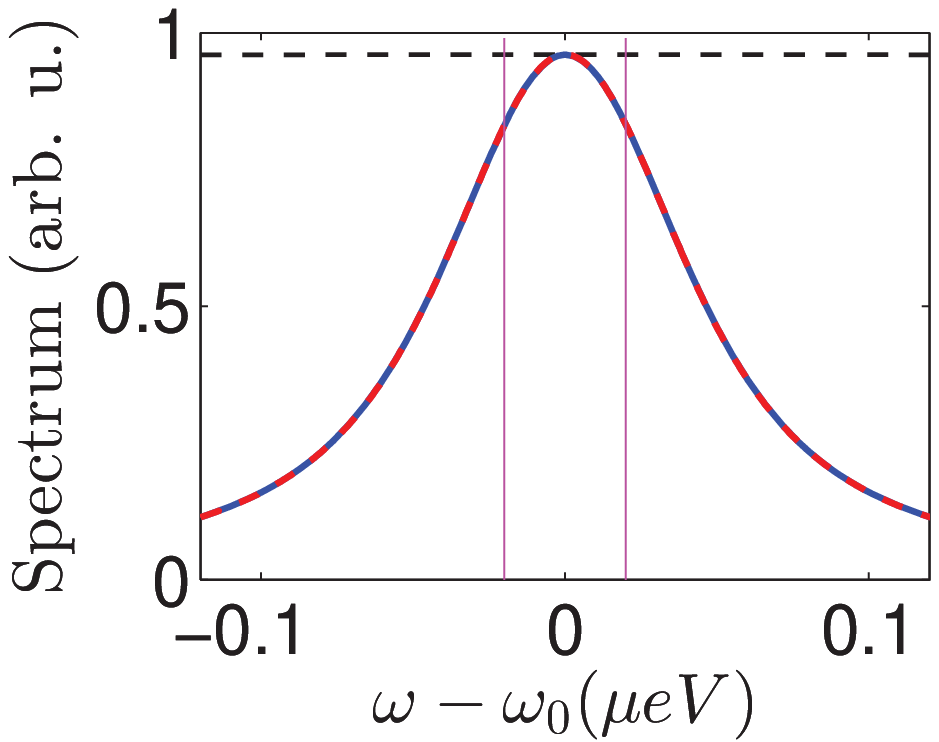}}
        \vspace{-0.1cm}
        \caption{\footnotesize{\red{Incoherent spectra (Mollow triplets) from exciton-$a$ (blue solid) and exciton-$b$ (red - dashed); the result for only 1 QD exciton is shown in the (black) thin dashed line. All results
are at the X point in a cavity, and the vertical magenta lines indicate the single exciton dressed-state resonances. 
(a) $Q=3000$ cavity with $\Omega_a = 10\,\mu$eV.
(b)  $Q=3000$ cavity with $\Omega_{a/b} = \pm 10\,\mu$eV.
(c) $Q=500$ cavity with $\Omega_a = 180\,\mu$eV.
(d) $Q=3000$ cavity with $\Omega_a = 0.2\,\mu$eV.
}}}
        \label{fig:Mollow}
        \vspace{-0.1cm}
\end{figure}

 \red{\it CW-pumped Mollow triplets, nonuplets and  singlets.}\ 
 One of the most striking experimental signatures of high-field cw driven
 two-level systems is the Mollow triplet~\cite{PhysRev.188.1969}, which results from transitions between the field-driven dressed states. Recently the Mollow triplet
in QD systems  has  been observed
 in a number of QD cavity systems~\cite{MollowQD1,MollowQD2,PhysRevLett.103.167402}. Using Eq.~(\ref{eq:LME}) and  the quantum regression theorem, 
the incoherent spectrum emitted from each QD
exciton, $n$, is obtained from~\cite{Carmichael1999}
$S_0^{n}(\omega) = \lim_{t\rightarrow\infty}\text{Re}[\int_0^{\infty}d\tau(\braket{\sigma_{ng}(t+\tau)
\sigma_{gn}(t)}
-\braket{\sigma_{ng}(t)}\braket{\sigma_{gn}(t)})e^{i(\omega_L-\omega)\tau}]$,
%\end{align}
where we assume the detector is aligned with the corresponding polarization and we ignore additional filtering effects associated with light propagation from the QD to the detector (though these effects
can  be included~\cite{PhysRevB.93.115308}).
We also consider the case where the QD excitons are directly pumped with an effective Rabi field, otherwise they will scale with $Q$
 and $n_g$ if pumped through the cavity mode and PC waveguide mode, respectively.

\red{Figures \ref{fig:Mollow}(a-b) show the Mollow
triplet for the cavity case above, with one and two coherent fields,
which demonstrate how the Mollow peaks are sharpened and clearly resolved, even though we are not in the Mollow regime ($\Omega\ll \Gamma_{a}$, cf. the broad black dashed spectrum from a single exciton.). Indeed, in the latter case, we have added a pure dephaing rate of 0.2 $\mu$eV, otherwise the peaks are infitesimally sharp. The next two panels show  examples of some striking physics: (c) shows how to observe more than three spectral peaks, as we are now dealing with a dressed triplet of states, which yields 9 resonances, 5 of which are degenerate, so 5 resolvable peaks can be seen in general; similar peaks have been predicted for V-type 3-level atom  when the dipole moments are nearly parallel~\cite{PhysRevLett.77.3995}.
While (d) demonstrates how to excite a single subnatural resonance, 
which has applications for producing single photon sources~\cite{PhysRevLett.108.093602}.}

%{\it Discussion.}\ \red{to do}

{\it Conclusions.}\ We have introduced several practical QD systems
that can yield substantial dipole-dipole coupling between orthogonal
dipoles within the same QD, through carefully nanoengineering the photonic AVI effects.
% This AVI coupling  can be used to coherently couple the orthogonal dipoles, %and in general such effects may also show up in a number of emerging experiments.
%For an antisymmetric entangled pair state, the dipole-dipole coupling completely %cancels radiative decay.
\red{We have also shown how to exploit such physics for generating a population trapping state and demonstrated the  consequences of these states for exploring high field physics, such  the Mollow triplet regime, with relatively weak fields. A wide range of other quantum optical effects should be accessible in this regime, including the possibility of exploring cavity-QED effects with cavities that are nominally in the weak coupling regime. 
Moreover, our formalism can easily be extended to multiple QDs, e.g., for use in  chiral spin networks \cite{PhysRevA.91.042116}. }

We thank Andrew Young and Ben Lang for useful discussions.
This work was funded by the Natural Sciences and Engineering Research Council of Canada (NSERC) and Queen's University.

% gives titles
\bibliographystyle{unsrt}
%\bibliographystyle{apalike}
%\bibliographystyle{plainnat}
%\bibliographystyle{aipauth4}
%\bibliographystyle{h-physrev}
%\bibliographystyle{apsrev4-1}
%\bibliographystyle{plainnt}
%\def\urlprefix{}
%\def\url#1{}

% URLd link with no titles
\bibliographystyle{aipnum4-1}
\bibliography{paperquantbib}

\end{document}